# Tunneling behavior of bismuth telluride nanoplates in electrical transport

Mustafa Eginligil [a], Weiqing Zhang [b], Alan Kalitsov [a], Xianmao Lu [b,*],

and Hyunsoo Yang [a,*]

[a]*Department of Electrical and Computer Engineering, National University of Singapore, 117576 Singapore*
[b]*Department of Chemical and Biomolecular Engineering, National University of Singapore, 117576 Singapore*

**Abstract.** We study the electrical transport properties of ensembles of bismuth telluride ($Bi_2Te_3$) nanoplates grown by solution based chemical synthesis. Devices consisting of $Bi_2Te_3$ nanoplates are fabricated by surface treatment after dropping the solution on the structured gold plates and the temperature dependence of resistance shows a nonmetallic behavior. Symmetric tunneling behavior in *I-V* was observed in both our experimental results and theoretical calculation of surface conductance based on a simple Hamiltonian, which excludes carrier-carrier interactions. Here, we present two devices: one showing symmetric, the other showing a two-step tunneling behavior. The latter can be understood in terms of disorder.

* E-mail address: chelxm@nus.edu.sg (X. Lu), eleyang@nus.edu.sg (H. Yang)

## 1. Introduction

Bismuth telluride ($Bi_2Te_3$), a semiconductor with an indirect bulk energy band gap of 0.165 eV [1], is a unique multifunctional material. It is an attractive thermoelectric material with the highest figure of merit ($ZT$ = 0.68) at room temperature in its bulk [2]. It was recently shown that in thin films of $Bi_2Te_3$ $ZT$ can be enhanced about ten times due to line dislocations in topologically protected perfectly conductive one dimensional state [3-5]. This physical property is under investigation, yet it is a well-known fact that $Bi_2Te_3$, like other members of its family (i.e. $Bi_2Se_3$ and $Sb_2Te_3$) exhibit exotic properties in low dimension, such topological insulating state with a single Dirac cone [6]. Dirac cone was clearly shown by angle resolved photoemisson spectroscopy (ARPES) studies [1], which have motivated a strong interest to elaborate dissipationless spin currents at room temperature in this material [7-10], in addition to the aforementioned superior thermoelectric properties.

In order to understand this newly discovered feature of $Bi_2Te_3$ family and make use of it for spintronics applications, there have been many electrical transport studies [11-19]. However, most of these observations have demonstrated that bulk transport dominates due to low conductivity of surface states compared to bulk and it appears as metallic behavior in resistance vs. temperature. There are some recent experimental studies which show an insulating behavior in electrical transport of $Bi_2Te_3$ family [11, 16, 19]. One of them attributes this insulating behavior to the coupling of top and bottom surfaces, as in conventional semiconductors [19]. Recently, scanning tunneling microscopy (STM) studies showed a high transmission probability of surface states which are topologically protected [20], making tunneling in transport of $Bi_2Te_3$ family a crucial exploratory field. However, tunneling has not been studied in detail for these materials.

It is well established that quantum spin Hall (QSH) state is robust against weak non-magnetic impurities and surface states are not affected, leading to high mobilities. However, recent work on the effect of strong disorder in HgTe/CdTe quantum wells, has shown the existence of Anderson insulator behavior [21]. Depending on the strength of disorder, the size of the samples, and the position of the Fermi level, it is possible to observe the quantization of conductance. It is also possible to observe Anderson insulator behavior in disordered $Bi_2Te_3$ family with strong spin-orbit interaction [22], in which disordered insulating bulk and conducting surface states coexist. This suggests $Bi_2Te_3$ family can be promising Anderson insulator candidates [23]. However, so far $Bi_2Te_3$ has been reported to show metallic behavior except the case where $Bi_2Te_3$ was prepared via a special cleaving after high temperature synthesis in a vacuum quartz tube [16].

Here, we present the current-voltage ($I$-$V$) characteristics of an ensemble of $Bi_2Te_3$ nanoplates (NPs) for temperatures between 6 K and 300 K which exhibit insulating behaviour. Hexagonal $Bi_2Te_3$ NPs with an average size of 250 nm and a thickness about 10 nm were prepared via a solution method. These plates were then dispersed in ethanol or chloroform and deposited onto gold pads of 100 nm thickness. Devices are formed, after the solution dries and leaving aggregates of the nanoplates on and between the gold contacts. In a device containing many NPs a clear asymmetric and two-step tunneling behavior appears at temperatures lower than 200 K, even though a simple theory based on the newly discovered state of matter, which does not account for carrier-carrier interactions, suggests symmetric *I-V* curves at all temperatures. Based on a recent theory [21], we discuss the possibility that this behavior may be due to randomly distributed impurities in $Bi_2Te_3$ NPs.

## 2. Chemical Synthesis and Structural Characterization

The chemical synthesis of $Bi_2Te_3$ NPs is as follows. In a glove box, 0.0386 g $Bi(CH_3COO)_3$ was mixed with 4 mL squalane. 0.5 M of Tellurium (Te) in trioctylphosphine (TOP) was prepared by dissolving 0.033 g of Te in 2.6 mL of TOP and stirring for two hours in the glove box. Then, 4 mL of $Bi(CH_3COO)_3$ solution in squalane was mixed with 0.7 mL of Te(TOP), followed by rapidly injection of the mixture into a three-neck flask equipped with a condenser under nitrogen protection. The reaction mixture was heated to 250 °C for 15 minutes. Black precipitate was isolated by centrifugation, followed by washing with chloroform, for five times. The resulting $Bi_2Te_3$ NPs were collected and dispersed in hexane.

The inset to Figure 1(d) shows a typical scanning electron microscopy (SEM) image of NPs. The NPs solution then was drop cast onto a 300 nm $Si/SiO_2$ substrate with lithographically patterned 100 nm thick gold plates. Upon evaporation of the solvent, aggregate of $Bi_2Te_3$ NPs formed between gold plates. The substrate was then subjected to a surface treatment with ozone stripping at 140 °C for 20 minutes in order to eliminate carbon radical residues. We perform X-ray photoemission spectroscopy (XPS) just before (red dashed curve) and after (black solid curve) the latter process to confirm the removal of the chemical residues, as in Figure 1(a) and 1(b), for Bi and Te peaks, respectively. In Figure 1(a), Bi $4f^{7/2}$ and $4f^{5/2}$ peaks at 160 eV and 165.3 eV, respectively appear after surface treatment; similarly in Figure 1(b) Te $3d^{5/2}$ and $3d^{3/2}$ peaks at 577.2 and 587.6 eV, respectively appear after surface treatment. In Figure 1(c), the X-ray diffraction (XRD) data of $Bi_2Te_3$ NPs show that it is single phase, and the peaks are indexed to the rhombohedral $Bi_2Te_3$. The calculated lattice parameters are $a$ = 4.369 Å and $c$ = 30.423 Å, slightly smaller than those of JCPDS 08-0021, $a$ = 4.381 Å and $c$ = 30.483 Å. After the surface treatment we measured Raman spectroscopy with excitation of 785 nm, with a power of ~ 0.2 mW, as seen in Figure 1(d). We observe the optical phonon mode $A_{1u}$ which is Raman-inactive in bulk, but infra-red active due to crystal symmetry breaking, as reported for mechanically exfoliated $Bi_2Te_3$ [24].

## 3. Electrical Transport Results and Discussion

The samples were annealed at 250 °C for 30 minutes to obtain electrical continuity between the gold plates and the ensemble of NPs. SEM images of two different junctions with separations of 2 µm and 5 µm for devices 1 and 2 are shown in Figure 2(a) and 2(b), respectively. Device 1 contains fewer NPs than device 2. The resistance as a function of temperature is measured by applying a current of 1 µA. For comparison, we amplified the resistance of device 2 by 35 times. The electrical transport is throughout many conduction paths between source and drain. The resistance of the device with more conduction paths would be lower. This qualitatively explains why device 2 has about 35 times less resistance. Both samples show an insulating behavior as indicated in Figure 2(c), with device 2 exhibiting a relatively rapid increase in resistance at temperatures lower than 20 K which may be due to carrier-carrier interactions [25]. For device 1, we observe symmetric tunneling I-V curves at all temperatures. As we increase the temperature from 6 K up to 300 K, the measured current increases for the same applied voltage, as shown in the *I-V* characteristics in Figure 3. This is in agreement with the resistance vs. temperature data in Figure 2(c).

This system can be simply thought of $Bi_2Te_3$ NPs in between two metal (M) contacts, i.e. M/$Bi_2Te_3$/M. We can make a qualitative comparison between experiment and theory by making use of a two dimensional system described by the Bernevig-Hughes-Zhang (BHZ) model, excluding carrier-carrier interactions [26]. The Hamiltonian which describes such a system is given as

$$H = \sum_{i,\sigma,\alpha} \varepsilon_\alpha c^+_{i\alpha\sigma} c_{i\alpha\sigma} - \sum_{ia\sigma\alpha\beta} t_{a\sigma} c^+_{i+a\alpha\sigma} c_{i\beta\sigma} , \qquad (1)$$

where $a$ label the four nearest neighbor sites, σ describes the spin, α, β is the orbital index, $c^+_{i\alpha\sigma}$ and $c_{i\alpha\sigma}$ are the creation and annihilation operators of the conduction electron on site $i$ with the orbital index $α$ and spin $σ$. The electron hopping integral is a 2×2 matrix in this basis

$$t_{a\sigma} = \begin{pmatrix} t_{ss} & t_{sp} e^{i\sigma\theta_a} \\ t_{sp} e^{-i\sigma\theta_a} & -t_{pp} \end{pmatrix}, \qquad (2)$$

where $θ_a$ is the angle of the nearest-neighbor bond $a$ with the $x$-axis. The bands near the Γ-point become inverted if

$$\varepsilon_s - \varepsilon_p \leq 4(t_{ss} + t_{pp}) . \qquad (3)$$

Our calculations of the *I-V* characteristics of M/$Bi_2Te_3$/M junctions with $N$=1 to 4 monolayer (ML) at room temperature are shown in Figure 4. The *I-V* curves strongly depend on the number of MLs. The current density for $N$=4 is much larger than that ones for $N$=1-3, where $N$ is the number of ML. On the other hand, there is no big difference between *I-V* curves for $N$=2 and $N$=3. This effect can be explained in terms of the different density of states for different number of MLs.

Independent of the thickness, there is a symmetric behavior of the *I-V* curves which stays symmetric as temperature decreases. However, it is not possible to make a direct quantitative comparison with experiment since NPs are quite randomly distributed. Ideally, it is expected that only surface states are conductive and the largest contribution to the measured current comes from the surface (as calculated in the simulation). In addition, the current density at the surface must be much larger while in the bulk it is very small. Therefore, the experimental current density is expected to be not uniform within the sample. Nevertheless, the symmetric tunneling behavior in *I-V* curves in the simulations is similar to our experimental observations in device 1 and there is a qualitative agreement. Extension of this simple model to 3D and inclusion of carrier-carrier interactions would provide more insight into tunneling studies and for understanding the behavior in device 2, as in the following.

For device 2, we measure asymmetry in *I-V* as shown in Figure 5 which is accompanied by a peak in the junction resistance versus bias voltage appearing around 0.5 V at 6 K as shown in Figure 6. As we increase the temperature gradually to 300 K, the asymmetry in *I-V* changes into symmetric *I-V* as in device 1. The inset to Figure 6 clearly reveals this behavior as we plot the calculated resistance (voltage/current) vs. bias voltage. This secondary hump has a resistance value of approximately 10 % (~45 kΩ) of the resistance value (~0.45 MΩ) around zero applied voltage. At 6 K, a potential barrier ~ 0.5 V gets activated which weakens as temperatures increases. We extract this potential barrier by Gaussian fitting of the calculated resistance vs. bias voltage data as shown in the inset of Figure 6.

Although the origin of this feature is not clear, we ascribe this temperature dependent activation potential to the existence of randomized impurities as predicted in Anderson insulators in which surface states are formed due to impurities induced in the system [21, 22]. In the case of strong disorder, at certain current densities, it is possible to observe local conductance extrema [22]. As temperature increases, this effect may be smeared out due to phonon scattering. Alternatively, it is possible that for positive applied voltages hopping mechanism is leading to a decrease in the current density compared to negative applied voltages in which the current density is exponentially dependent on the applied voltage. Although this could be valid for samples with relatively lower resistance (e.g. device 2), it might not be possible to resolve for samples with higher resistance (e.g. device 1). Recently, we observed similar I-V characteristics in $Bi_2Te_3$ flakes that are mechanically exfoliated, i.e. two step tunneling behavior at much lower positive applied voltages.

## 4. Conclusion

In conclusion, the *I-V* characteristics of ensemble of chemically synthesized $Bi_2Te_3$ nanoplates have shown a tunneling behavior as determined from *I-V* curves measured from 6 K up to room temperature. As expected from a simple 2D Hamiltonian considering no carrier-carrier interactions, a symmetric tunneling behavior has been observed, except one showing a two-step tunneling behavior. The latter is believed to be due to randomly distributed impurities, but it is also possible due to unconventional hopping mechanism. Future theoretical elaborations and experimental investigations on disorder and transport on single nanoplates could give insight on this promising thermoelectric and spintronic material for device applications.

**Acknowledgement**

This work is supported by the Singapore National Research Foundation under CRP Award No. NRF-CRP 4-2008-06.

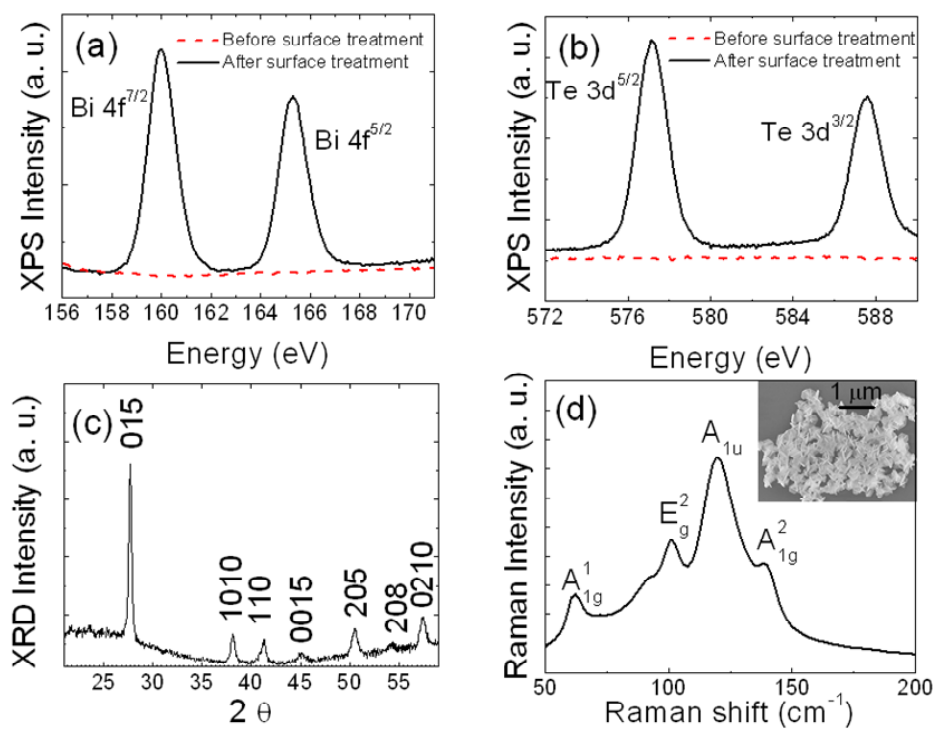

**FIGURE 1** The XPS before and after the surface treatment for Bi (a) and for Te (b). (c) The XRD results show single phase. (d) Raman spectroscopy of an ensemble of $Bi_2Te_3$ nanoplates, with 785 nm excitation. $A_{1u}$ mode is observed around 119 cm$^{-1}$. The inset shows the SEM image of the ensemble of mostly hexagonal $Bi_2Te_3$ nanoplates with a size of 250 nm by solution based chemical synthesis.

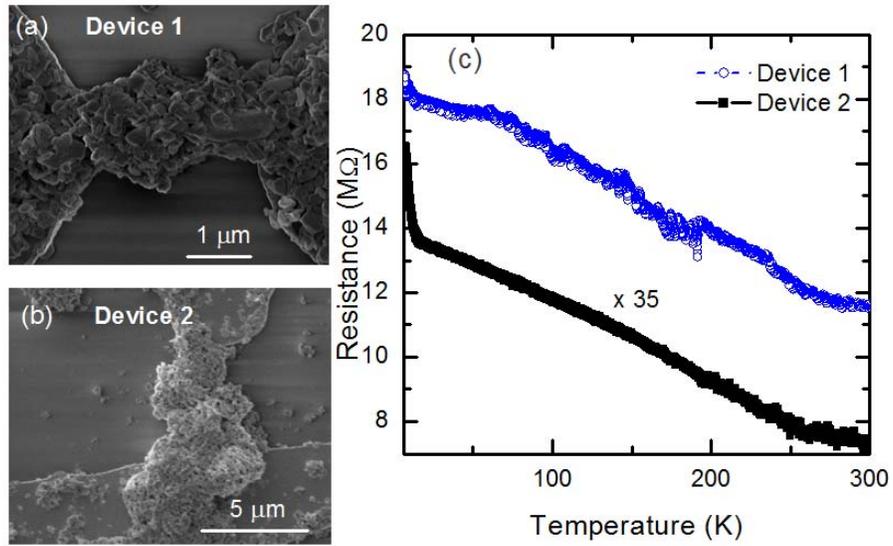

**FIGURE 2** The SEM images of device 1 (a) and device 2 (b) of ensembles of $Bi_2Te_3$ nanoplates, with 2 μm and 5 μm separation of gold contact plates, respectively. Resistance vs. temperature at 1 μA for device 1 (open symbol) and device 2 (filled symbol) are shown in (c) in which the resistance of device 2 is amplified 35 times in order to compare the two samples.

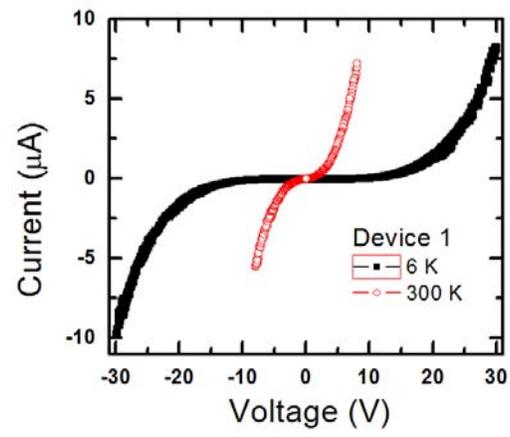

**FIGURE 3** The symmetric tunneling behavior of device 1 at 6 K (filled square) and 300K (open circle).

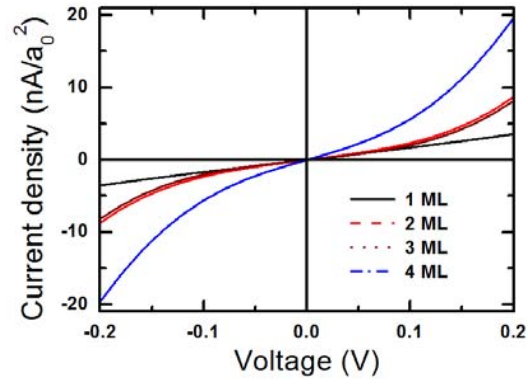

**FIGURE 4** The simulation results of *I-V* curves at room temperature for 1 – 4 ML of the simple model based on the theory excluding carrier-carrier interactions and disorder shows the symmetric behavior.

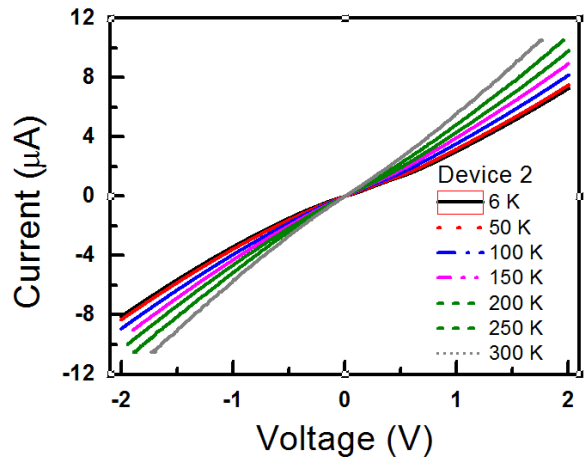

**FIGURE 5** The *I-V* characteristics of device 2, in which temperature is increased from 6 K to 300K. Asymmetry with respect to zero voltage becomes much clear as temperature decreases.

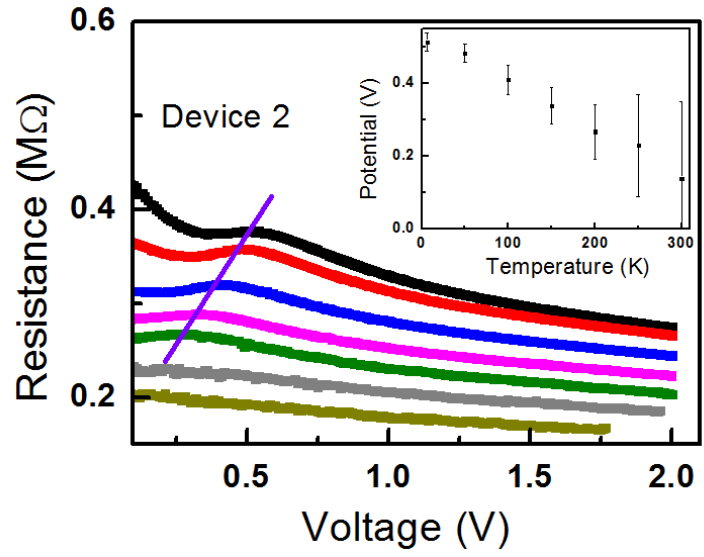

**FIGURE 6** The plot of calculated resistance vs. bias voltage in which a clear hump is observed at 0.5 V at the lowest temperature (6 K). It weakens in magnitude and the potential drops as the temperature increases which is negligible above 200 K. The inset is the plot of the potential vs. temperature as determined from Gaussian fittings.